# A new Eulerian theory of turbulence constrained by random Galilean invariance

R. V. R. Pandya*

*Department of Mechanical Engineering,*

*University of Puerto Rico at Mayaguez,*

*Mayaguez, Puerto Rico, PR 00681, USA*

## Abstract

We propose a new Eulerian turbulence theory to obtain a closed set of equations for homogeneous, isotropic turbulent velocity field correlations and propagator functions by incorporating constraints of random Galilean invariance. This incorporation generates a few different equations for propagator and the present theory suggests a way to utilize them into the closure solutions of two-time and single-time velocity correlations' equations so as to properly account for random sweeping phenomena. The present theory yields *exact* solutions when applied to simple model problem of random oscillator.

# INTRODUCTION

An attempt to solve closure problem of fluid turbulence led Kraichnan to propose direct interaction approximation (DIA) [1, 2] as a pioneer renormalized perturbation theory (RPT), followed by other RPTs [3–22] which have been reviewed from time to time [19, 23–26].

The DIA closed set of equations includes equations governing the evolution of ensemble averaged response function, single-time and two-time velocity correlations. The energetically consistent DIA [27] has failed to be consistent with the result of Kolmogorov's similarity theory [28, 29] which has been accepted as a first crucial test for any closure scheme [23, 24]. Th failure occurs due to the presence of divergence in the response function equation [23]. Kraichnan attributed the failure to Eulerian framework and not to DIA. Kraichnan applied DIA and related higher order approximations to an idealized convection problem and showed that Eulerian framework is unsuited for these approximations to maintain exactly the symmetry of convective invariance of single-time triple correlation of velocity field [30]. Later, this symmetry was displayed in sharp form as a fundamental property of the exact dynamics - invariance under random Galilean transformation (RGT) [31] - and which is violated by DIA. Consequently, to overcome the failure, Kraichnan proposed a generalized Lagrangian velocity field [30] and implemented DIA in this Lagrangian framework which resulted in a theory known as Lagrangian history direct interaction (LHDI) [31]. The LHDI is partly compatible with all the transformation rules as provided by RGT [31].

Here it is salutary to mention that at present LHDI does not retain a basic property of the governing Navier-Stokes equation and Lagrangian field that the Eulerian quantities can be calculated independent of the Lagrangian quantities. A need to retain the basic property would require improving the LHDI and/or formulating the closure theory in the Eulerian framework only. In this paper, an attempt is made to formulate a new Eulerian turbulence theory constrained by RGT and which would find applications to other closure problems which do not have any Lagrangian representation. Before presenting the new theory, we now briefly mention a few facts of another relevant Eulerian theory, namely local energy transfer (LET) theory of McComb [24].

Based on the Edwards theory [4], the LET was proposed by McComb [32] in an Eulerian framework. Since then, LET has evolved into a closed set of equations comprising equations governing the evolution of single-time and two-time velocity correlations and fluctuation-dissipation ansatz governing the propagator function [15, 33]. The LET is compatible with Kolmogorov spectrum [24] despite its failure to comply with RGT. The LET has been remained under persistent surveillance, especially of McComb and co-workers, for its performance and accomplishments in cases of isotropic turbulence and related passive scalar convection [33–38]. Within the framework of LET, yet another closed set of equations were suggested and named as variant of local energy transfer (VLET) [39]. The VLET and LET differ in the form for propagator equation. So far it is unclear why one should use VLET propagator equation instead of LET fluctuation-dissipation ansatz? The present theory also provides justification for VLET propagator equation and suggests limiting its use only to the closure equation for single-time velocity correlation.

Now we present basic equations and set of closure equations of DIA, LET and VLET. Then we present the new Eulerian theory,its application to random oscillator problem and concluding remarks.

## BASIC EQUATIONS AND DIA, LET, VLET CLOSURE EQUATIONS

### Navier-Stokes equation and closure problem

Consider a homogeneous, isotropic, incompressible fluid turbulence in a reference frame $S$ which is stationary in the laboratory. The Eulerian velocity field $u_i(\mathbf{x}, t)$ in physical space-time $(\mathbf{x}, t)$ can be expressed in terms of Fourier modes of the velocity field $u_i(\mathbf{k}, t)$ by using the Fourier transform

$$u_i(\mathbf{x}, t) = \int d^3\mathbf{k}\, u_i(\mathbf{k}, t) \exp(\iota \mathbf{k}.\mathbf{x}), \tag{1}$$

where $\iota = \sqrt{-1}$ and $\mathbf{k}$ is the wave vector and subscripts take the value 1, 2 or 3. The Fourier transform of Navier-Stokes equations governing the dynamics of $u_i(\mathbf{x}, t)$ can be written as [23, 24]

$$\left(\frac{\partial}{\partial t} + \nu k^2\right) u_i(\mathbf{k}, t) = M_{ijm}(\mathbf{k}) \int d^3\mathbf{p}\, u_j(\mathbf{p}, t) u_m(\mathbf{k} - \mathbf{p}, t), \tag{2}$$

and the continuity equation is

$$k_j u_j(\mathbf{k}, t) = 0. \tag{3}$$

Here $\nu$ is kinematic viscosity of the fluid, inertial transfer operator $M_{ijm}(\mathbf{k})$ is given by

$$M_{ijm}(\mathbf{k}) = (2\iota)^{-1}\left[k_j P_{im}(\mathbf{k}) + k_m P_{ij}(\mathbf{k})\right] \tag{4}$$

and by using Kronecker delta $\delta_{ij}$, the projector $P_{ij}(\mathbf{k})$ can be written as

$$P_{ij}(\mathbf{k}) = \delta_{ij} - k_i k_j |\mathbf{k}|^{-2}. \tag{5}$$

Denoting the operation of taking an ensemble average by $\langle\ \rangle$, equations governing the evolution of two-time velocity correlation

$$Q_{in}(\mathbf{k},-\mathbf{k};t,t') = \langle u_i(\mathbf{k},t) u_n(-\mathbf{k},t') \rangle \tag{6}$$

and single-time velocity correlation

$$Q_{in}(\mathbf{k},-\mathbf{k};t,t) = \langle u_i(\mathbf{k},t) u_n(-\mathbf{k},t) \rangle \tag{7}$$

can be obtained from Eq. (2) and written as

$$\left(\frac{\partial}{\partial t} + \nu k^2\right) Q_{in}(\mathbf{k},-\mathbf{k};t,t') = M_{ijm}(\mathbf{k}) \int d^3\mathbf{p}\, \langle u_j(\mathbf{p},t) u_m(\mathbf{k}-\mathbf{p},t) u_n(-\mathbf{k},t') \rangle, \tag{8}$$

and

$$\left(\frac{\partial}{\partial t} + 2\nu k^2\right) Q_{in}(\mathbf{k},-\mathbf{k};t,t) = M_{ijm}(\mathbf{k}) \int d^3\mathbf{p}\, \langle u_j(\mathbf{p},t) u_m(\mathbf{k}-\mathbf{p},t) u_n(-\mathbf{k},t) \rangle$$
$$+ M_{njm}(-\mathbf{k}) \int d^3\mathbf{p}\, \langle u_j(\mathbf{p},t) u_m(-\mathbf{k}-\mathbf{p},t) u_i(\mathbf{k},t) \rangle. \tag{9}$$

For homogeneous, isotropic turbulence these correlations can be further written as

$$Q_{in}(\mathbf{k},\mathbf{k}';t,t') = P_{in}(\mathbf{k}) Q(k;t,t') \delta(\mathbf{k}+\mathbf{k}'), \tag{10}$$

$$Q_{in}(\mathbf{k},\mathbf{k}';t,t) = P_{in}(\mathbf{k}) Q(k;t,t) \delta(\mathbf{k}+\mathbf{k}'), \tag{11}$$

where $\delta$ represents the Dirac delta function. These Eqs. (8) and (9) pose well known closure problem of turbulence due to the presence of unknown third-order velocity correlations in them.

To solve the closure problem, Kraichnan also utilized equation for infinitesimal response tensor $\hat{G}_{in}(\mathbf{k};t,t')$ for Eq. (2) in his DIA, which can be written as [23, 24]

$$\left(\frac{\partial}{\partial t} + \nu k^2\right) \hat{G}_{in}(\mathbf{k};t,t') - 2M_{ijm}(\mathbf{k}) \int d^3\mathbf{p}\, u_j(\mathbf{p},t) \hat{G}_{mn}(\mathbf{k}-\mathbf{p};t,t') = P_{in}(\mathbf{k}) \delta(t-t'). \tag{12}$$

Now we present closure solution as obtained in DIA and LET theories alongwith closed set of equations suggested as VLET.

**DIA, LET and VLET equations**

The DIA yields closed set of equations for $Q_{in}(\mathbf{k}, -\mathbf{k}; t, t')$, $Q_{in}(\mathbf{k}, -\mathbf{k}; t, t)$ and for ensemble averaged response tensor $G_{in}(\mathbf{k}; t, t') = \langle \hat{G}_{in}(\mathbf{k}; t, t') \rangle$. Whereas LET and VLET provides closed set of equations for $Q_{in}(\mathbf{k}, -\mathbf{k}; t, t')$, $Q_{in}(\mathbf{k}, -\mathbf{k}; t, t)$ and propagator $H_{in}(\mathbf{k}; t, t')$. In this section, we will use notation $H_{in}(\mathbf{k}; t, t')$ instead of notation $G_{in}(\mathbf{k}; t, t')$ in DIA equations with an understanding that it represents ensemble averaged response function when used in DIA and propagator when used in LET and VLET. Also for homogeneous, isotropic turbulence we have

$$H_{in}(\mathbf{k}; t, t') = P_{in}(\mathbf{k}) H(k; t, t'). \tag{13}$$

With this change of notation, it is easy to see that equations for $Q_{in}(\mathbf{k}, -\mathbf{k}; t, t')$ and $Q_{in}(\mathbf{k}, -\mathbf{k}; t, t)$ are identical in form in DIA, LET and VLET and for homogeneous, isotropic turbulence these can be written as [24, 39]

$$\left(\frac{\partial}{\partial t} + \nu k^2\right) Q(k; t, t') = P(k; t, t') \tag{14}$$

$$\left(\frac{\partial}{\partial t} + 2\nu k^2\right) Q(k; t, t) = 2P(k; t, t) \tag{15}$$

where the inertial transfer term $P(k; t, t')$ is

$$P(k; t, t') = \int d^3\mathbf{p}\, L(\mathbf{k}, \mathbf{p}) \Big[ \int_0^{t'} ds\, H(k; t', s) Q(p; t, s) Q(|\mathbf{k} - \mathbf{p}|; t, s)$$
$$- \int_0^t ds\, H(p; t, s) Q(k; t', s) Q(|\mathbf{k} - \mathbf{p}|; t, s) \Big] \tag{16}$$

and

$$L(\mathbf{k}, \mathbf{p}) = \frac{[\mu(k^2 + p^2) - kp(1 + 2\mu^2)](1 - \mu^2)kp}{k^2 + p^2 - 2kp\mu} \tag{17}$$

with $\mu$ represnting the cosine of the angle between the vectors $\mathbf{k}$ and $\mathbf{p}$. The equations for $H(k; t, t')\ \forall t > t'$ in DIA, LET and VLET are different in form and can be written as [24, 39]

$$\text{DIA:}\ \left(\frac{\partial}{\partial t} + \nu k^2\right) H(k; t, t') = -\int d^3\mathbf{p}\, L(\mathbf{k}, \mathbf{p}) \Big[ \int_{t'}^t ds$$
$$H(k; s, t') H(p; t, s) Q(|\mathbf{k} - \mathbf{p}|; t, s) \Big] + \delta(t - t') \tag{18}$$

$$\text{LET:}\ Q(k; t, t') = H(k; t, t') Q(k; t', t'), \tag{19}$$

$$\text{VLET:}\ Q(k; t, t) = H(k; t, t') Q(k; t, t'). \tag{20}$$

The LET Eq. (19) represents the generalized fluctuation-dissipation relation.

Kraichnan derived Eq. (18) from Eq. (12) using RPT. The DIA fails to capture Kolmogorov spectrum due the divergence present in the Eq. (18) [23]. Whereas McComb's LET theory does not utilize Eq. (12) at all to arrive at fluctuation-dissipation ansatz i.e. Eq. (19). The LET equations are compatible with the Kolmogorov spectrum and the Eq. (19) does not contain any divergence. It should be noted that VLET was suggested within the framework of LET. Also, VLET equation for propagator is different than the LET propagator equation and so far there is no analysis available to decide which one is more appropriate. In this respect, the present new theory also provide more insight into proper use of propagator Eq. (20) and a few other equations for propagators generated in the present theory. In fact, later we will observe that VLET Eq. (20) is obtained from propagator equations which properly account for random sweeping phenomena appearing in the equation for $Q(k;t,t')$. Next we present the new Eulerian theory of turbulence.

**NEW RANDOM GALILEAN EULERIAN THEORY (RGET)**

In the present Eulerian theory, hereafter referred to as random Galilean Eulerian theory (RGET), Navier-Stokes (NS) equation is first transformed by using Kraichnan's concept of random Galilean invariance [30, 31]. Then the closure is solved starting from the transformed NS equation, hereafter referred to as random Galilean Navier-Stokes (RGNS) equation.

Now we present RGNS equation. In reference frame $S$, we add a uniform Galilean velocity field $v_i$ to a realization of $u_i(\mathbf{k}, t)$ and obtain a homogeneous turbulent field $u_i(\mathbf{k}, t)_v$ governed by

$$\left(\frac{\partial}{\partial t} + \nu k^2\right) u_i(\mathbf{k}, t)_v = -\iota k_j v_j u_i(\mathbf{k}, t)_v + M_{ijm}(\mathbf{k}) \int d^3\mathbf{p}\, u_j(\mathbf{p}, t)_v u_m(\mathbf{k} - \mathbf{p}, t)_v. \tag{21}$$

Following Kraichnan [31], consider $v_i$ to be statistically distributed over the ensemble of infinite realizations with $\langle v_j \rangle = 0$ and statistically independent of $u_i(\mathbf{k}, t)$. Also, assume Gaussian distribution for $v_i$ for later simplicity with

$$v_0^2 = \langle v_1^2 \rangle = \langle v_2^2 \rangle = \langle v_3^2 \rangle. \tag{22}$$

The Eq. (21) with random Galilean velocity $v_i$ is referred to as RGNS equation. Also, the continuity equation is

$$k_j u_j(\mathbf{k}, t)_v = 0. \tag{23}$$

From Eq. (21), the governing equations for two-time and single-time velocity correlations of velocity field $u_i(\mathbf{k}, t)_v$ can be written as

$$\left(\frac{\partial}{\partial t} + \nu k^2\right) Q_{in}(\mathbf{k}, -\mathbf{k}; t, t')_v = \langle -\iota k_j v_j u_i(\mathbf{k}, t)_v u_n(-\mathbf{k}, t')_v \rangle$$
$$+ M_{ijm}(\mathbf{k}) \int \mathrm{d}^3\mathbf{p} \, \langle u_j(\mathbf{p}, t)_v u_m(\mathbf{k} - \mathbf{p}, t)_v u_n(-\mathbf{k}, t')_v \rangle, \quad (24)$$

and

$$\left(\frac{\partial}{\partial t} + 2\nu k^2\right) Q_{in}(\mathbf{k}, -\mathbf{k}; t, t)_v = M_{ijm}(\mathbf{k}) \int \mathrm{d}^3\mathbf{p} \, \langle u_j(\mathbf{p}, t)_v u_m(\mathbf{k} - \mathbf{p}, t)_v u_n(-\mathbf{k}, t)_v \rangle$$
$$+ M_{njm}(-\mathbf{k}) \int \mathrm{d}^3\mathbf{p} \, \langle u_j(\mathbf{p}, t)_v u_m(-\mathbf{k} - \mathbf{p}, t)_v u_i(\mathbf{k}, t)_v \rangle. \quad (25)$$

It should be noted that $v_j$ containing term does not appear explicitly in the Eq. (25) for single-time velocity correlation. For homogeneous, isotropic turbulence these correlations can be further written as

$$Q_{in}(\mathbf{k}, \mathbf{k}'; t, t')_v = P_{in}(\mathbf{k}) Q(k; t, t')_v \delta(\mathbf{k} + \mathbf{k}'), \tag{26}$$

$$Q_{in}(\mathbf{k}, \mathbf{k}'; t, t)_v = P_{in}(\mathbf{k}) Q(k; t, t)_v \delta(\mathbf{k} + \mathbf{k}'). \tag{27}$$

It is easy to see that solutions of Eqs. (2) and (21) are related by

$$u_i(\mathbf{k}, t)_v = \exp\left[-\iota k_j v_j (t - t_0)\right] u_i(\mathbf{k}, t) \tag{28}$$

and from which follow the exact random Galilean transformation (RGT) rules [30, 31] for statistical quantities of physical interests, written as

$$Q(k; t, t')_v = \exp\left[-\frac{k^2 v_0^2}{2}(t - t')^2\right] Q(k; t, t'), \tag{29}$$

$$\langle -\iota k_j v_j u_i(\mathbf{k}, t)_v u_n(-\mathbf{k}, t')_v \rangle = -k^2 v_0^2 (t - t') Q_{in}(\mathbf{k}, -\mathbf{k}; t, t')_v, \tag{30}$$

$$\langle u_j(\mathbf{p}, t)_v u_m(\mathbf{k} - \mathbf{p}, t)_v u_n(-\mathbf{k}, t')_v \rangle = \exp\left[-\frac{k^2 v_0^2}{2}(t - t')^2\right] \langle u_j(\mathbf{p}, t) u_m(\mathbf{k} - \mathbf{p}, t) u_n(-\mathbf{k}, t') \rangle. \tag{31}$$

It should be noted that the random Galilean velocity transformation does not affect single-time velocity correlations and we have

$$Q(k; t, t)_v = Q(k; t, t), \tag{32}$$

$$\langle u_j(\mathbf{p}, t)_v u_m(\mathbf{k} - \mathbf{p}, t)_v u_n(-\mathbf{k}, t)_v \rangle = \langle u_j(\mathbf{p}, t) u_m(\mathbf{k} - \mathbf{p}, t) u_n(-\mathbf{k}, t) \rangle. \tag{33}$$

Now we perform the following steps:

1. Use DIA type of renormalized perturbation (RP) method to obtain closure solutions for random sweeping term $\langle -\iota k_j v_j u_i(\mathbf{k},t)_v u_n(-\mathbf{k},t')_v \rangle$. Obtain the constraints for which RP closure solution for the random sweeping term becomes identical to *exact* solution given by Eq. (30). We should mention here that these constraints yield governing equations for propagator and DIA type derivation of response function equation is not required in the present theory.

2. Obtain RP closure solution for two-time triple velocity correlation $\langle u_j(\mathbf{p},t)_v u_m(\mathbf{k}-\mathbf{p},t)_v u_n(-\mathbf{k},t')_v \rangle$ appearing on the right hand side (rhs) of Eq. (24) . Use the constraint equations for propagator appropriately into RP closure solution for two-time triple velocity correlation. Use $v_0 = 0$ into Eq. (24) to obtain final equation for $Q(k;t,t')$.

3. The Eq. (25) does not contain any term with $v_0$, the step similar to the first step cannot be performed to obtain equation for propagator. In such a case, we rely on obtained propagator equations in the first step for $Q(k;t,t')_v$ equation. We substitute $t = t'$ in those propagator equations to obtain propagator appearing in the closure solution for single-time triple velocity correlations. We must mention that this kind of obtained propagator yields *exact* solution for single-time correlation in case of analogous simple model problem of random oscillator and relevant details are presented later in this paper. Use of all these along with $v_0 = 0$ into Eq. (25) then yield final equation for $Q(k;t,t)$.

Now we provide details of these steps to obtain closure solution.

**Step 1**

By using the usual recipe of DIA [23, 24], renormalized perturbation closure solution for random sweeping term can be obtained from RGNS equation (21) and can be written as

$$\langle -\iota k_j v_j u_i(\mathbf{k},t)_v u_n(-\mathbf{k},t')_v \rangle = -k^2 v_0^2 \int_0^t \mathrm{d}s\, H_{ia}(\mathbf{k};t,s)_v Q_{an}(\mathbf{k},-\mathbf{k};s,t')_v$$
$$+ k^2 v_0^2 \int_0^{t'} \mathrm{d}s\, H_{na}(-\mathbf{k};t',s)_v Q_{ai}(\mathbf{k},-\mathbf{k};t,s)_v, \qquad (34)$$

where $H_{ia}(\mathbf{k};t,s)_v$ is propagator for RGNS equation. Though the renormalized perturbation equation for $H_{ia}(\mathbf{k};t,s)_v$ can be derived similar to DIA, here we follow a different approach to obtain equations for propagator. It should be noted that rhs of Eq. (34) differs from the *exact* solution given by Eq. (30). The rhs of Eq. (34) can be made identical to the exact solution if we substitute in the first integral ($\int_0^t$)

$$H_{ia}(\mathbf{k};t,s)_v Q_{an}(\mathbf{k},-\mathbf{k};s,t')_v = Q_{in}(\mathbf{k},-\mathbf{k};t,t')_v \tag{35}$$

and substitute in the second integral ($\int_0^{t'}$)

$$H_{na}(-\mathbf{k};t',s)_v Q_{ai}(\mathbf{k},-\mathbf{k};t,s)_v = Q_{in}(\mathbf{k},-\mathbf{k};t,t')_v. \tag{36}$$

For homogeneous, isotropic turbulence these can be further written as

$$Q(k;t,t')_v = H(k;t,s)_v Q(k;s,t')_v, \tag{37}$$

$$Q(k;t,t')_v = H(k;t',s)_v Q(k;t,s)_v. \tag{38}$$

These two constraint equations (35) and (36) are two different equations for propagator and exactly capture random sweeping phenomena appearing in integrals ($\int_0^t$) and ($\int_0^{t'}$), respectively. In view of this, following rules should be used later to properly capture sweeping effects in the RP closure solution of second term on the rhs of Eq. (24).

- Use Eq. (35) for propagator which appears in terms containing integral ($\int_0^t$) in the RP closure solution of second term on the rhs of, Eq. (24).

- Use Eq. (36) for propagator which appears in terms containing integral ($\int_0^{t'}$) in the RP closure solution of second term on the rhs of, Eq. (24).

**Step 2**

By using the usual recipe of DIA [23, 24], RP closure solution for the second term on the rhs of Eq. (24) can be obtained. Further using Eqs. (34), (35) and (36) and the above-mentioned rules, the RP closure equation for $Q(k;t,t')$ for homogeneous, isotropic turbulence can be written as

$$\left(\frac{\partial}{\partial t} + \nu k^2\right) Q(k;t,t')_v = -k^2 v_0^2 (t-t') Q(k;t,t')_v + P_1(k,t,t')_v + P_2(k,t,t')_v, \tag{39}$$

where

$$P_1(k,t,t')_v = -\int d^3\mathbf{p}\, L(\mathbf{k},\mathbf{p}) \int_0^t ds\, H_1(p;t,s)_v Q(k;t',s)_v Q(|\mathbf{k}-\mathbf{p}|;t,s)_v, \tag{40}$$

$$H_1(p,t,s)_v Q(p,s,t')_v = Q(p,t,t')_v \tag{41}$$

$$P_2(k,t,t')_v = \int d^3\mathbf{p}\, L(\mathbf{k},\mathbf{p}) \int_0^{t'} ds\, H_2(k;t',s)_v Q(p;t,s)_v Q(|\mathbf{k}-\mathbf{p}|;t,s)_v, \tag{42}$$

$$H_2(k,t',s)_v Q(k;t,s)_v = Q(k,t,t')_v. \tag{43}$$

It should be noted that we have introduced $H_1$ and $H_2$ to distinguish propagators governed by Eqs. (37) and (38), respectively.

The Eq. (39) can be simplified further by substituting for $H_1$, $H_2$ and along with $v_0 = 0$ yield the final RGET renormalized perturbation equation for $Q(k;t,t')$, written as

$$\left(\frac{\partial}{\partial t} + \nu k^2\right) Q(k;t,t') = -\int d^3\mathbf{p}\, L(\mathbf{k},\mathbf{p}) Q(p;t,t') \int_0^t ds\, \frac{1}{Q(p,s,t')} Q(k;t',s) Q(|\mathbf{k}-\mathbf{p}|;t,s)$$
$$+ \int d^3\mathbf{p}\, L(\mathbf{k},\mathbf{p}) Q(k;t,t') \int_0^{t'} ds\, \frac{1}{Q(k;t,s)} Q(p;t,s) Q(|\mathbf{k}-\mathbf{p}|;t,s) \tag{44}$$

**Step 3:**

Substituting $t = t'$ into Eqs. (35) and (36) yields identical equations for propagator, written as

$$H_{ia}(\mathbf{k};t,s)_v Q_{an}(\mathbf{k},-\mathbf{k};s,t)_v = Q_{in}(\mathbf{k},-\mathbf{k};t,t)_v \tag{45}$$

and for homogeneous, isotropic turbulence which reduces to

$$H(k;t,s)_v Q(k;s,t)_v = Q(k;t,t)_v. \tag{46}$$

It should be noted that for $v_0 = 0$, Eq. (46) becomes identical to the VLET propagator Eq. (20).

Now, the DIA type of RP closure solution of Eq. (25) for homogeneous, isotropic turbulence can be written as

$$\left(\frac{\partial}{\partial t} + 2\nu k^2\right) Q(k;t,t)_v = 2\int d^3\mathbf{p}\, L(\mathbf{k},\mathbf{p}) \left[ -\int_0^t ds\, H(p;t,s)_v Q(k;t,s)_v Q(|\mathbf{k}-\mathbf{p}|;t,s)_v \right.$$
$$\left. + \int_0^t ds\, H(k;t,s)_v Q(p;t,s)_v Q(|\mathbf{k}-\mathbf{p}|;t,s)_v \right] \tag{47}$$

where the propagators $H(p;t,s)_v$ and $H(k;t,s)_v$ appearing in this equation are governed by the constraint equation (46). Substituting Eq. (46) alongwith $v_0 = 0$ into Eq. (47) yield the final RGET renormalized perturbation equation for $Q(k;t,t)$, which can be written as

$$\left(\frac{\partial}{\partial t} + 2\nu k^2\right) Q(k;t,t) = -2\int d^3\mathbf{p}\, L(\mathbf{k},\mathbf{p}) Q(p;t,t) \int_0^t ds\, \frac{Q(k;t,s)}{Q(p,t,s)} Q(|\mathbf{k}-\mathbf{p}|;t,s)$$
$$+2\int d^3\mathbf{p}\, L(\mathbf{k},\mathbf{p}) Q(k;t,t) \int_0^t ds\, \frac{Q(p;t,s)}{Q(k;t,s)} Q(|\mathbf{k}-\mathbf{p}|;t,s). \tag{48}$$

These two equations (44) and (48) form the closure solution of homogeneous, isotropic turbulence as obtained by the present random Galilean Eulerian theory (RGET).

It should be noted that the present theory do not need RP closure equation for Kraichnan's response function $G_{in}(\mathbf{k}; t, t')$. But to show the application of various above mentioned steps, we now provide details of closure solution for $G_{in}(\mathbf{k}; t, t')$ equation as obtained by the RGET.

**RGET CLOSURE SOLUTION FOR KRAICHNAN'S RESPONSE FUNCTION**

The ensemble averaged response tensor $G_{in}(\mathbf{k}; t, t')_v = \langle \hat{G}(\mathbf{k}; t, t')_v \rangle$ of RGNS equation (21) is governed by

$$\left(\frac{\partial}{\partial t} + \nu k^2\right) G_{in}(\mathbf{k}; t, t')_v = -\langle \iota k_j v_j \hat{G}_{in}(\mathbf{k}; t, t')_v \rangle + 2M_{ijm}(\mathbf{k}) \Big[ \int d^3\mathbf{p} \, \langle u_j(\mathbf{p}, t)_v \hat{G}_{mn}(\mathbf{k} - \mathbf{p}; t, t')_v \rangle \Big] + P_{in}(\mathbf{k})\delta(t - t') \quad (49)$$

The exact solution for the first term on the rhs is

$$-\langle \iota k_j v_j \hat{G}_{in}(\mathbf{k}; t, t')_v \rangle = -k^2 v_0^2 (t - t') G_{in}(\mathbf{k}; t, t')_v. \quad (50)$$

The DIA type of RP closure solution within the framework of present theory is

$$-\langle \iota k_j v_j \hat{G}_{in}(\mathbf{k}; t, t')_v \rangle = -k^2 v_0^2 \int_{t'}^{t} ds \, H_{ia}(\mathbf{k}; t, s)_v G_{an}(\mathbf{k}, s, t')_v \quad (51)$$

and which becomes identical to the exact solution given by Eq. (50) under the constraint

$$G_{in}(\mathbf{k}; t, t')_v = H_{ia}(\mathbf{k}; t, s)_v G_{an}(\mathbf{k}, s, t')_v. \quad (52)$$

This, for homogeneous, isotropic turbulence can be written as

$$G(k; t, t')_v = H(k; t, s)_v G(k; s, t')_v. \quad (53)$$

It should be noted that $H_{ia}(\mathbf{k}; t, s)_v$ is a propagator for Eq. (49) and is obtained here by invoking the constraint.

Now within the present theory framework, the RP closure solution of Eq. (49) for homogeneous, isotropic turbulence can be written as

$$\left(\frac{\partial}{\partial t} + \nu k^2\right) G(k; t, t')_v = -k^2 v_0^2 (t - t') G(k; t, t')_v - \int d^3\mathbf{p} \, L(\mathbf{k}, \mathbf{p}) \Big[ \int_{t'}^{t} ds \, H(p; t, s)_v G(k; s, t')_v Q(|\mathbf{k} - \mathbf{p}|; t, s)_v \Big] + \delta(t - t') \quad (54)$$

where $H(p;t,s)_v$ is governed by equation similar to Eq. (53), i.e.

$$G(p;t,t')_v = H(p;t,s)_v G(p;s,t')_v. \tag{55}$$

Substituting it into Eq. (54) along with $v_0 = 0$ result into closure equation for Kraichnan's response function for NS equation as derived by the RGET. The final equation can be written as

$$\left(\frac{\partial}{\partial t} + \nu k^2\right) G(k;t,t') = -\int d^3\mathbf{p}\, L(\mathbf{k},\mathbf{p}) G(p;t,t') \int_{t'}^{t} ds\, \frac{G(k;s,t')}{G(p;s,t')} Q(|\mathbf{k}-\mathbf{p}|;t,s) + \delta(t-t'). \tag{56}$$

It should be noted that this RGET equation (56) differs from the Kraichnan's DIA response function Eq. (18).

**APPLICATION OF PRESENT THEORY TO RANDOM OSCILLATOR PROBLEM**

**Exact solutions of random oscillator problem**

The amplitude of a random oscillator is governed by the equation [23]

$$\left(\frac{d}{dt} + \nu + \iota b\right) q(t) = 0, \tag{57}$$

where $\nu$ is a real damping parameter and $b$ is a real time-independent parameter which is statistically distributed over the ensemble of infinite realizations of the oscillator. For later simplicity, we assume Gaussian distribution for $b$. The infinitesimal response function $\hat{G}(t,t')$ for Eq. (57) is governed by

$$\left(\frac{d}{dt} + \nu + \iota b\right) \hat{G}(t,t') = 0 \ \forall t > t' \tag{58}$$

and $\hat{G}(t',t') = 1$. The exact solution for Eqs. (57) and (58), with initial value $q(0)$, can be written as

$$q(t) = q(0) \exp[-\nu t - \iota bt] \tag{59}$$

and

$$\hat{G}(t,t') = \exp[-\nu(t-t') - \iota b(t-t')]. \tag{60}$$

Using Eqs. (59) and (60), exact two-time and single-time second-order correlation for $q(t)$ and ensemble averaged response function $H(t,t')$ can be obtained and written as

$$Q(t,t') = \langle q(t)q(t')\rangle = Q(0,0) \exp\left[-\nu(t+t') - \frac{\langle b^2 \rangle}{2}(t+t')^2\right] \tag{61}$$

$$Q(t,t) = \langle q(t)q(t) \rangle = Q(0,0)\exp[-2\nu t - 2\langle b^2 \rangle t^2] \tag{62}$$

$$G(t,t') = \langle \hat{G}(t,t') \rangle = \exp\left[-\nu(t-t') - \frac{\langle b^2 \rangle}{2}(t-t')^2\right] \tag{63}$$

**The present theory solutions**

It should be noted that the term $\iota b q(t)$ in Eq. (57) is analogous to the term containing $v_i$ in RGNS equation (21). So here we use various propagator equations already obtained in the present theory directly for the solution of Eq. (57) without including any random Galilean velocity type parameter. And if their use obtain *exact* solutions for the random oscillator, that will further justify the propagator equations. In fact, we do obtain exact solutions and details of which are present next.

Using DIA type RP theory, the closure equation for $Q(t,t')$ can be obtained from Eq. (eq:16) and written as

$$\left(\frac{d}{dt} + \nu\right) Q(t,t') = -\langle b^2 \rangle \left[\int_0^t ds\, H(t,s)Q(t',s) + \int_0^{t'} ds\, H(t',s)Q(t,s)\right]. \tag{64}$$

Now according to the present theory, we substitute propagator equations

$$Q(t,t') = H(t,s)Q(t',s) \tag{65}$$

and

$$Q(t,t') = H(t',s)Q(t,s), \tag{66}$$

which are identical in form to Eqs. (37) and (38), into the terms containing integral ($\int_0^t$) and integral ($\int_0^{t'}$), respectively. This yields

$$\left(\frac{d}{dt} + \nu\right) Q(t,t') = -\langle b^2 \rangle (t+t') Q(t,t') \tag{67}$$

and whose solution is identical to the exact solution given by Eq. (61).

In case of $Q(t,t)$, using RP theory we obtain

$$\left(\frac{d}{dt} + 2\nu\right) Q(t,t) = -4\langle b^2 \rangle \int_0^t ds\, H(t,s)Q(t,s) \tag{68}$$

and using

$$Q(t,t) = H(t,s)Q(t,s) \tag{69}$$

into Eq. (68) yields exact solution given by Eq. (62). The propagator equation (69) is identical in form to the Eq. (46) and it justifies the use of Eq. (46) in case of single-time turbulent velocity correlation.

The RP closure solution for $G(t, t')$ can be written as

$$\left(\frac{d}{dt} + \nu\right) G(t, t') = -\langle b^2 \rangle \int_{t'}^{t} ds\, H(t, s) G(s, t') \tag{70}$$

and using equation $G(t, t') = H(t, s) G(s, t')$ which is identical in form to Eq. (53), we obtain exact solution for $G(t, t')$ as given by Eq. (63).

## CONCLUDING REMARKS

In this paper, we have solved closure problem of turbulence by utilizing a newly proposed random Galilean Eulerian theory (RGET) of turbulence. In this theory, the starting equation is random Galilean Navier-Stokes (RGNS) equation (21) instead of Navier-Stokes equation (2). To solve the closure problems involved in equations for statistical properties of RGNS equation, first a DIA type of renormalized perturbation (RP) method is utilized and then the constraints of random Galilean invariance transformation are invoked in RP equations. These constraints yield different equations for propagators appearing in different integral terms of obtained RP equations. The RGET provides certain rules to utilize these propagators properly and it has become possible to eliminate the propagators from the closed set of RP equations for $Q(k; t, t')_v$ and $Q(k; t, t)_v$. Also, the application of RGET to a simple model problem of random oscillator has resulted in an exact solution for the involved closure problem.

Here we should mention that, within the framework of RGET, one can select any suitable renormalized perturbation method instead of DIA type but we have preferred DIA. Other choices remain to be explored. The assessment of RGET against numerical solution of NS equation and application of RGET to other related closure problems, e.g. passive scalar and differential diffusion of passive scalars, will be considered in future work.

―――――――

* raja.pandya(AT)upr.edu, rvrptur(AT)yahoo.com

[1] R. H. Kraichnan, Phys. Rev. **109**, 1407 (1958).

[2] R. H. Kraichnan, J. Fluid Mech. **5**, 497 (1959).

[3] H. W. Wyld Jr., Ann. Phys. **14**, 143 (1961).

[4] S. F. Edwards, J. Fluid Mech. **18**, 239 (1964).

[5] L. L. Lee, Ann. Phys. **32**, 292 (1965).

[6] J. R. Herring, Phys. Fluids **8**, 2219 (1965).

[7] J. R. Herring, Phys. Fluids **9**, 2106 (1966).

[8] S. F. Edwards and W. D. McComb, J. Phys. A **2**, 157 (1969).

[9] R. Phythian, J. Phys. A **2**, 181 (1969).

[10] R. Balescu and A. Senatorski, Ann. Phys. **58**, 587 (1970).

[11] S. A. Orszag, J. Fluid Mech. **41**, 363 (1970).

[12] T. Nakano, Ann. Phys. **73**, 326 (1972).

[13] T. Nakano, Phys. Fluids. **31**, 1420 (1988).

[14] P. C. Martin, E. D. Siggia, and H. A. Rose, Phys. Rev. A **8**, 423 (1973).

[15] W. D. McComb, J. Phys. A **11**, 613 (1978).

[16] H. Horner and R. Lipowsky, Z. Physik B **33**, 223 (1979).

[17] Y. Kaneda, J. Fluid Mech. **107**, 131 (1981).

[18] J. Qian, Phys. Fluids **26**, 2098 (1983).

[19] V. S. L'vov, Phys. Rep. **207**, 1 (1991).

[20] C. Mou and P. B. Weichman, Phys. Rev. E **52**, 3738 (1995).

[21] V. S. L'vov and I. Procaccia, Phys. Rev. E. **52**, 3840 (1995).

[22] V. S. L'vov and I. Procaccia, Phys. Rev. E. **52**, 3858 (1995).

[23] D. C. Leslie, *Developments in the Theory of Turbulence* (Clarendon Press, Oxford, 1973).

[24] W. D. McComb, *The Physics of Fluid Turbulence* (Oxford University Press, New York, NY, 1990).

[25] W. D. McComb, Rep. Prog. Phys. **58**, 1117 (1995).

[26] M. Lesieur, *Turbulence in Fluids*, 3rd ed. (Kluwer, Dordrecht, 1997).

[27] R. H. Kraichnan, J. Math. Phys. **2**, 124 (1961).

[28] A. N. Kolmogorov, Dokl. Akad. Nauk SSSR **30**, 301 (1941).

[29] G. K. Batchelor, *The Theory of Homogeneous Turbulence* (Cambridge University Press, Cambridge, UK, 1959).

[30] R. H. Kraichnan, Phys. Fluids **7**, 1723 (1964).

[31] R. H. Kraichnan, Phys. Fluids **8**, 575 (1965).

[32] W. D. McComb, J. Phys. A **7**, 632 (1974).

[33] W. D. McComb, M. J. Filipiak, and V. Shanmugasundaram, J. Fluid Mech. **245**, 279 (1992).

[34] W. D. McComb and V. Shanmugasundaram, J. Fluid Mech. **143**, 95 (1984).

[35] W. D. McComb, V. Shanmugasundaram, and P. Hutchinson, J. Fluid Mech. **208**, 91 (1989).

[36] W. D. McComb and A. P. Quinn, Physica A **317**, 487 (2003).

[37] J. S. Frederiksen, A. G. Davies, and R. C. Bell, Phys. Fluids **6**, 3153 (1994).

[38] J. S. Frederiksen and A. G. Davies, Geophys. Astrophys. Fluid Dyn. **92**, 197 (2000).

[39] R. V. R. Pandya, Phys. Rev. E **70**, 066307 (2004).